# 30 inch Roll-Based Production of High-Quality Graphene Films for Flexible Transparent Electrodes

Sukang Bae[1*], Hyeong Keun Kim[3*], Youngbin Lee[1], Xianfang Xu[5], Jae-Sung Park[7], Yi Zheng[5], Jayakumar Balakrishnan[5], Danho Im[2], Tian Lei[1], Young Il Song[6], Young Jin Kim[1,3], Kwang S. Kim[7], Barbaros Özyilmaz[5], Jong-Hyun Ahn[1,4†], Byung Hee Hong[1,2†] & Sumio Iijima[1,8]

[1]SKKU Advanced Institute of Nanotechnology (SAINT) and Center for Human Interface Nano Technology (HINT), [2]Department of Chemistry, [3]Department of Mechanical Engineering, [4]School of Advanced Materials Science and Engineering, Sungkyunkwan University, Suwon 440-746, Korea. [5]NanoCore & Department of Physics, National University of Singapore, Singapore 117576 & 117542, [6]Digital & IT Solution Division, Samsung Techwin, Seongnam 462-807, Korea. [7]Center for Superfunctional Materials, Department of Chemistry, Pohang University of Science and Technology, Hyojadong, Namgu, Pohang 790-784, Korea,[8]Nanotube Research Center, National Institute of Advanced Industrial Science and Technology (AIST), Tsukuba 305-8565 & Faculty of Science and Engineering, Meijo University, Nagoya 468-8502, Japan

*These authors contributed equally to this work.

†To whom correspondence should be addressed. E-mail: byunghee@skku.edu or ahnj@skku.edu

**We report that 30-inch scale multiple roll-to-roll transfer and wet chemical doping considerably enhance the electrical properties of the graphene films grown on roll-type Cu substrates by chemical vapor deposition. The resulting graphene films shows a sheet resistance as low as ~30 Ω/sq at ~90 % transparency which is superior to commercial transparent electrodes such as indium tin oxides (ITO). The monolayer of graphene shows sheet resistances as low as ~125 Ω/sq with 97.4% optical transmittance and half-integer quantum Hall effect, indicating the**

**high-quality of these graphene films. As a practical application, we also fabricated a touch screen panel device based on the graphene transparent electrodes, showing extraordinary mechanical and electrical performances.**

Graphene and related materials have attracted tremendous attention for the last few years due to their fascinating electrical (*1*), mechanical (*2,3*), and chemical (*4,5*) properties. There have been many efforts to utilize these outstanding properties of graphene for macroscopic applications such as transparent conducting films useful for flexible/stretchable electronics (*6-8*). The implementation of graphene will require a production worthy process and recently X. Li et al (*11*) discovered and demonstrated a chemical vapor deposition process to grow graphene (monolayer graphite) on arbitrarily large Cu substrates (foils). Conventional transparent electrode, indium tin oxide (ITO), that is commonly used in solar cells, touch sensors and flat panel displays show a sheet resistance smaller than 100 Ohm/sq with ~90 % optical transparency as well as unlimited scalability, while the best reported sheet resistance for single-layer graphene is around ~350 Ohm/square (*6,9-11,20*).

   We have taken advantage of the CVD process to grow very large graphene films on Cu at high temperatures close to 1000 °C followed by a roll-based layer-by-layer transfer onto flexible substrates that includes an etching process to remove the metal catalyst layers which are obstacles for the direct use of graphene on as-grown substrates (*6, 9-10*). Therefore, the transfer of graphene films onto a foreign substrate is essential (*12,13*).The obstacle of graphene growth on rigid substrates has been overcome by the use of large flexible Cu foils (*11*), which enables the use of a roll-type substrate fitting the tubular shape of the furnace can maximize the scale and the homogeneity of graphene films. The flexibility of graphene and Cu foils (11) further allows efficient etching and transfer processes employing a cost and time-effective roll-to-roll production methods.

There are three essential steps in the roll-to-roll transfer (*14-16*) (Fig. 1A), which are i) adhesion of polymer supports to the graphene on the Cu foil, ii) etching of Cu layers, and iii) release of graphene layers and transfer on to a target substrate. In the adhesion step, the graphene film grown on a Cu foil is attached to a thin polymer support such as thermal-release tapes between two rollers. In the subsequent step, the Cu layers are removed by electrochemical reaction with a particular Cu etchant (*17, 18*). Finally, the graphene films are transferred from the polymer support onto a target substrate by removing the adhesive force on the polymer support. In the case of using thermal release tapes (*12, 13*), the graphene films are detached from the tapes and released to counter substrates by thermal treatment (Fig. 1A).

Fig. 2A-C show the photographs of roll-based synthesis and transfer processes. An 8-inch wide tubular quartz reactor is employed in the CVD system, where monolayer graphene films can be synthesized on a roll of Cu foil as large as 30-inch in diagonal direction (Fig. 2A). Usually, there exists a temperature gradient depending on the radial position inside a tubular reactor. This sometimes resulted in inhomogeneous growth of graphene on Cu foils in our preliminary work. To solve this problem, a ~7.5-inch quartz tube wrapped with a Cu foil was inserted and suspended inside the 8-inch quartz tube. Thus, the radial inhomogeneity in reaction temperature can be minimized (*18*). In the first step of synthesis, the roll of Cu foil is inserted to a tubular quartz tube and then heated up to 1000°C with flowing 10 sccm $H_2$ at 180 mTorr. After reaching 1000°C, the sample is annealed for 30 min without changing flow rate and pressure. As has been pointed out by Li et al, annealing of Cu can dramatically increase the grain size (*11*). Cu foils were heat treated to increase grain size from a few µm to the ~100 µm size, and we find that the larger grain size Cu foils yield higher quality graphene films (*11,18*). The gas mixture of $CH_4$ and $H_2$ is then flowed at 1.6 Torr with a rate of 30 sccm and 10 sccm for 15 min, respectively. Finally, the sample is rapidly cooled down to room temperature (~10°C/sec) with flowing $H_2$ under the pressure of 180 mTorr.

After the growth, the graphene film grown on the Cu foil is attached to a thermal release tape (Fine Chemical Co. and Nitto Denko Co.) by applying soft pressure (~0.2 MPa) between two rollers. After etching the Cu foil in a plastic bath filled with Cu etchant, the transferred graphene film on the tape is rinsed with DI water to remove residual etchant, and it is ready to be transferred to any kinds of flat or curved surfaces on demand. Subsequently, the graphene film on the thermal release tape is inserted to the rolls together with a target substrate and exposed to mild heat of 90~120°C for 3~5 min, resulting in the transfer of graphene films from the tape to the target substrate (Fig. 2B). By repeating these steps on the same substrate, multilayered graphene films can be prepared, showing enhanced electrical and optical properties as demonstrated by Li *et al.* using wet-transfer methods at centimeter scale (*20*). Fig. 2C shows the 30-inch multilayer graphene film transferred to a roll of 130 µm thick polyethylene terephthalate (PET) substrates. The scalability and the processability of CVD graphene and the roll-to-roll methods presented here are expected to enable the continuous production of graphene films in large scale.

Fig. 2D shows a screen-printing process to fabricate 4-wire touch screen panels (*19*) based on the graphene/PET transparent conducting films. After printing electrodes and dot spacers, the upper and lower panels are carefully assembled and connected to a controller installed in a lap top computer (Fig. 2E), which shows extraordinary electromechanical performances as we will discuss later (Fig. 2F) (*18*).

The graphene films look dominantly monolayers in the Raman spectra (Fig. 3A). However, atomic force microscope (AFM) and transmission electron microscope images often show bilayer and multilayer islands (*18*). As we transfer the graphene layers one after another (*20*), the intensities of G and 2D band peaks are increasing together, but their ratios don't change significantly. This is because the hexagonal lattices of upper and lower layers are randomly oriented unlike graphite so that the original properties of each monolayer remain unchanged even after staking into multilayers (*21,22*), which is clearly different from the case of multilayer graphene exfoliated from graphite crystals (*23*). The randomly stacked layers behave independently without significant change in electronic band structures, and the overall conductivity of graphene films appears to be proportional to the number of stacked layers (*20*). The optical transmittance is usually decreased by 2.2~2.3 % for an additional transfer, implying that the average thickness is approximately a monolayer (*24*).

The unique electronic band structure of graphene allows the modulation of charge carrier concentrations depending on electric field induced by gate bias (*25*) or chemical doping (*26*), resulting in enhancement of sheet resistance. We tried various types of chemical doping methods, and found that nitric acid ($HNO_3$) is very effective for the p-doping of graphene films. Fig. 3C shows Raman spectra of the graphene films before and after doping with 66 wt% $HNO_3$ for 5 min. The large peak shift ($\Delta v=18$ cm$^{-1}$) indicates that the graphene film is strongly p-doped. The shifted G peak is often split near the randomly stacked bilayer islands as shown in Fig. 3C. We suppose that the lower graphene layer screened by top layers experiences less doping effect, leading to the G-band splitting. In the X-ray photoelectron spectra (XPS), the C1s peaks corresponding to $sp^2$ and $sp^3$ hybridized states are shifted to lower energy, similar to the case of *p*-doped carbon nanotubes (*26*). On the other hand, multilayer stacking results in blue-shifted C1s peaks. We suppose that weak chemical bonding such as π–π stacking interaction causes descreening of nucleus charges, leading to the overall increases core electron binding energies. We also find that the work functions of graphene films estimated by ultraviolet photoelectron spectroscopy (UPS) are blue-shifted by ~130 meV with increasing doping time (Fig. 3D, inset), which would be very important to control the efficiency of photovoltaic (*27*) or light-emitting devices based on graphene transparent electrodes (*28*).

The electrical properties of graphene films formed through layer-by-layer staking methods are investigated. Usually, the sheet resistance of the graphene film with ~97% transmittance is as low as ~125 Ω/sq when it is transferred by a soluble polymer support such as polymethyl-methacrylate (PMMA) (*9, 10, 20*). However, the transferrable size of the wet transfer methods is limited below a few centimetres because of the weak mechanical strength of spin-coated PMMA layers, while the scale of roll-to-roll dry transfer assisted by a thermal release tape is in principle unlimited. In the process of roll-to-roll dry transfer, the first layer sometimes shows 2~3 times larger sheet resistance than the case of the PMMA-assisted wet transfer method. As the number of layers increases, the resistance drops faster compared to the wet transfer method (Fig. 4A). We suppose that the adhesion of the first layer with the substrate is not strong enough for the complete separation of graphene films from thermal release tapes. As a result, there can be mechanical damages on graphene films, leading to the increase of overall sheet resistance. Since additional layers are not directly affected by the adhesion with substrate surface, the sheet resistance of multilayers prepared by the roll-to-roll method doesn't differ much from the wet transfer case. The p-doping with HNO$_3$ clearly enhances the electrical properties of graphene films, and it is more effective for roll-to-roll processed graphene films. The sheet resistance of the *p*-doped 4-layer graphene film with ~90% optical transmittance is as low as ~30 Ω/sq, which is superior to common transparent electrodes such as ITO (*29*).

Standard e-beam lithography has been used to fabricate graphene hall bars on conventional 300nm Si/ SiO$_2$ (Fig. 4C). The left inset of Fig. 4C shows the four terminal resistance of such samples as a function of back-gate voltage ($V_{bg}$) at both room temperature in black and at low temperature (T = 7 K) and zero magnetic field. We observe the graphene specific gate bias dependence of the resistance with a sharp Dirac peak and a hall effect mobility of 7350 cm$^2$/Vs at low temperatures. This allows the observation of the quantum Hall effect at 6 K and a magnetic field of *B*=9T (Fig. 4C, right). The fingerprint of single layer graphene, the half-integer quantum Hall effect is observed with plateaus at filling factor *v* =2, 6, and 10 at $R_{xy}$ = 1/2, 1/6 and 1/10($h/e^2$), respectively. While the sequence of the plateaus remains intake for both the electron side and the hole side, there is a slight deviation from the fully quantized values on the electron side. We attribute this to the presence of grain boundaries. Finally, the electromechanical properties of graphene/PET touch screen panels are tested (Fig. 4D). Unlike an ITO-based touch panel that easily breaks under 1~2 % strain, the graphene-based panel stands up to 5% strain, which is limited not by graphene itself but by printed silver electrodes (Fig. 4D) (*30*).

In summary, we have developed and demonstrated a roll-to-roll process of graphene on ultra large Cu substrates. The multiple transfer and the simple chemical doping of graphene films considerably enhance the electrical/optical properties. Considering the outstanding scalability/processibility of roll-to-roll and CVD methods and the extraordinary flexibility/conductivity of graphene films, we expect that commercial production and application for large-scale transparent electrodes replacing the use of ITO can be realized in near future.

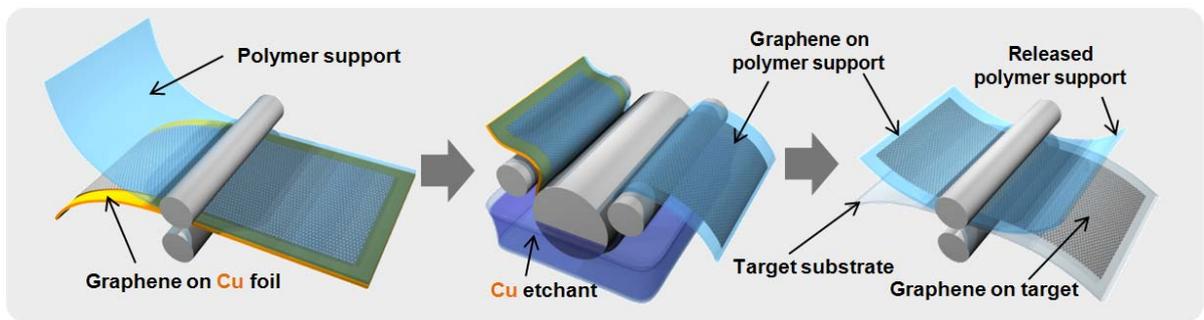

**Fig. 1**. Schematic of the roll-based production of graphene films grown on a Cu foil, including adhesion of polymer supports, Cu etching (rinsing), and dry transfer-printing on a target substrate. A wet chemical doping can be carried out using the similar set-up used for etching.

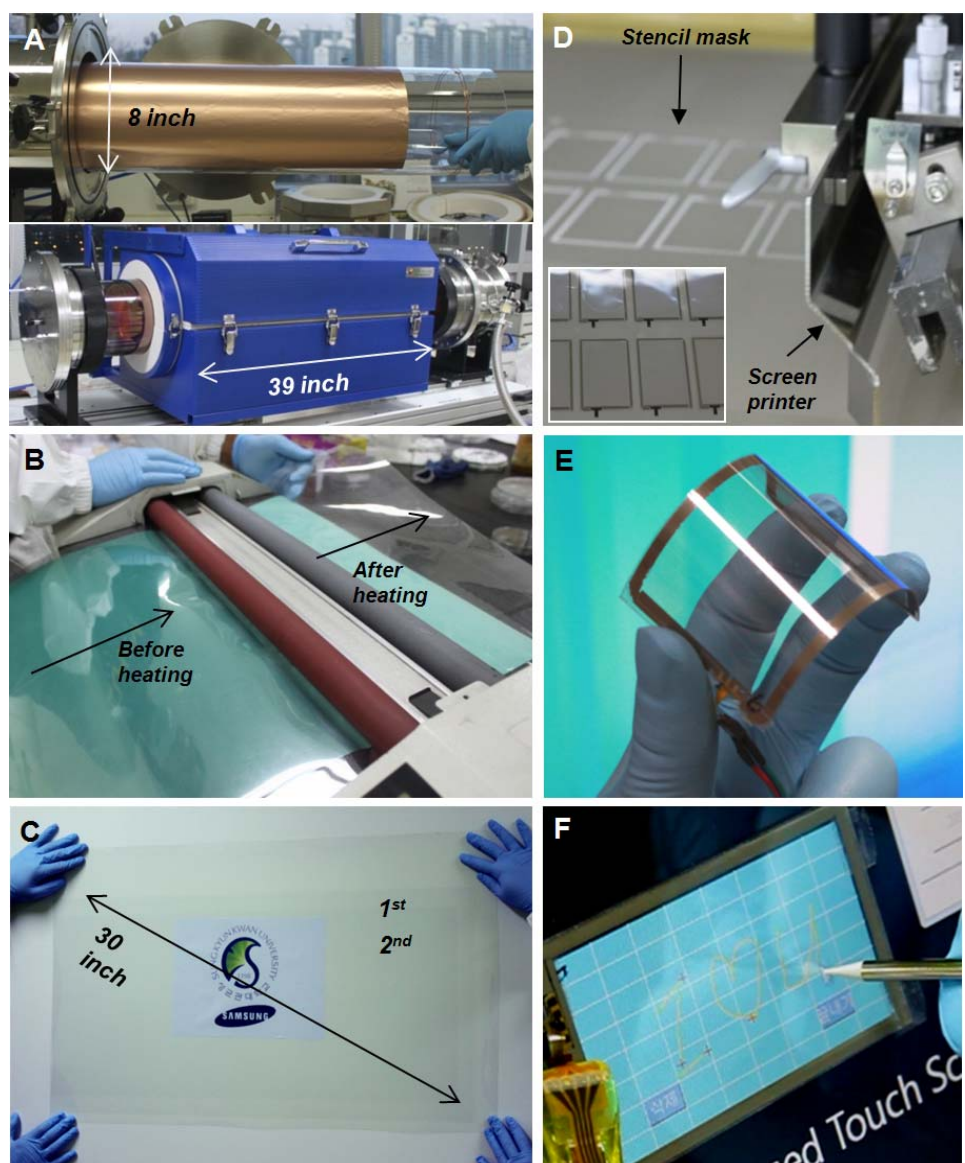

**Fig. 2**. Photographs of the roll-based production of graphene films. (**A**) A Cu foil wrapping around a 7.5–inch quartz tube to be inserted into an 8-inch quartz reactor. The lower image shows the Cu foil reacting with $CH_4$ and $H_2$ gases at high temperatures. (**B**) Roll-to-roll transfer of graphene films from a thermal release tape to a PET film at 120°C. (**C**) A transparent ultra-large-area graphene film transferred on a 35-inch PET sheet. (**D**) Screen printing process of silver paste electrodes on graphene/PET film. The inset shows 3.1-inch graphene/PET panels patterned with silver electrodes before assembly. (**E**) An assembled graphene/PET touch panel showing outstanding flexibility. (**D**) A graphene-based touch screen panel connected to a computer with control software. The operation movie is provided as an online supporting material.

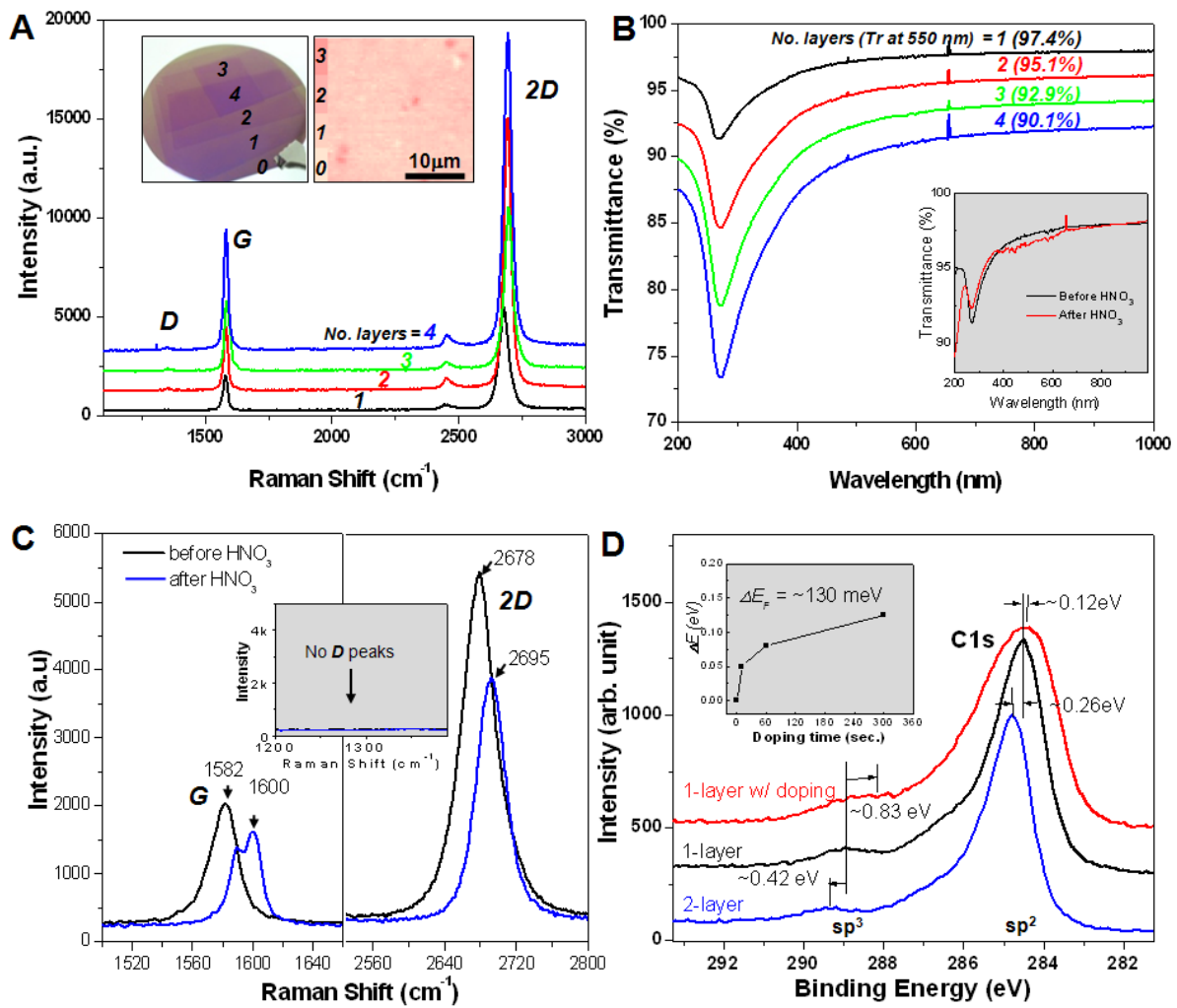

**Fig. 3**. Optical characterizations of the graphene films prepared by layer-by-layer transfer on $SiO_2$/Si and on PET substrates. (**A**) Raman spectra of graphene films with different number of stacked layers. The left inset shows a photograph of transferred graphene layers on a 4-inch $SiO_2$(300nm)/Si wafer. The right inset is a typical optical microscope image of the monolayer graphene, showing > 95% monolayer coverage. A PMMA-assisted transfer method is used for this sample. (**B**) UV-Visible spectra of roll-to-roll layer-by-layer transferred graphene films on PET substrates. The inset shows the UV spectra of graphene films with and without $HNO_3$ doping. (**C**) Raman spectra of $HNO_3$-doped graphene films, showing ~18 cm$^{-1}$ blue shift both for G and 2D peaks. D band peaks are not observed before and after doping, indicating that $HNO_3$ treatment is not destructive to the chemical bonds of graphene. (**D**) X-ray photoelectron spectroscopy (XPS) peaks showing typical red-shift and broadening caused by *p*-doping. Multilayer stacking shows a high-energy shift. The inset shows work function changes with respect to doping time, measure by ultraviolet photoelectron spectroscopy (UPS).

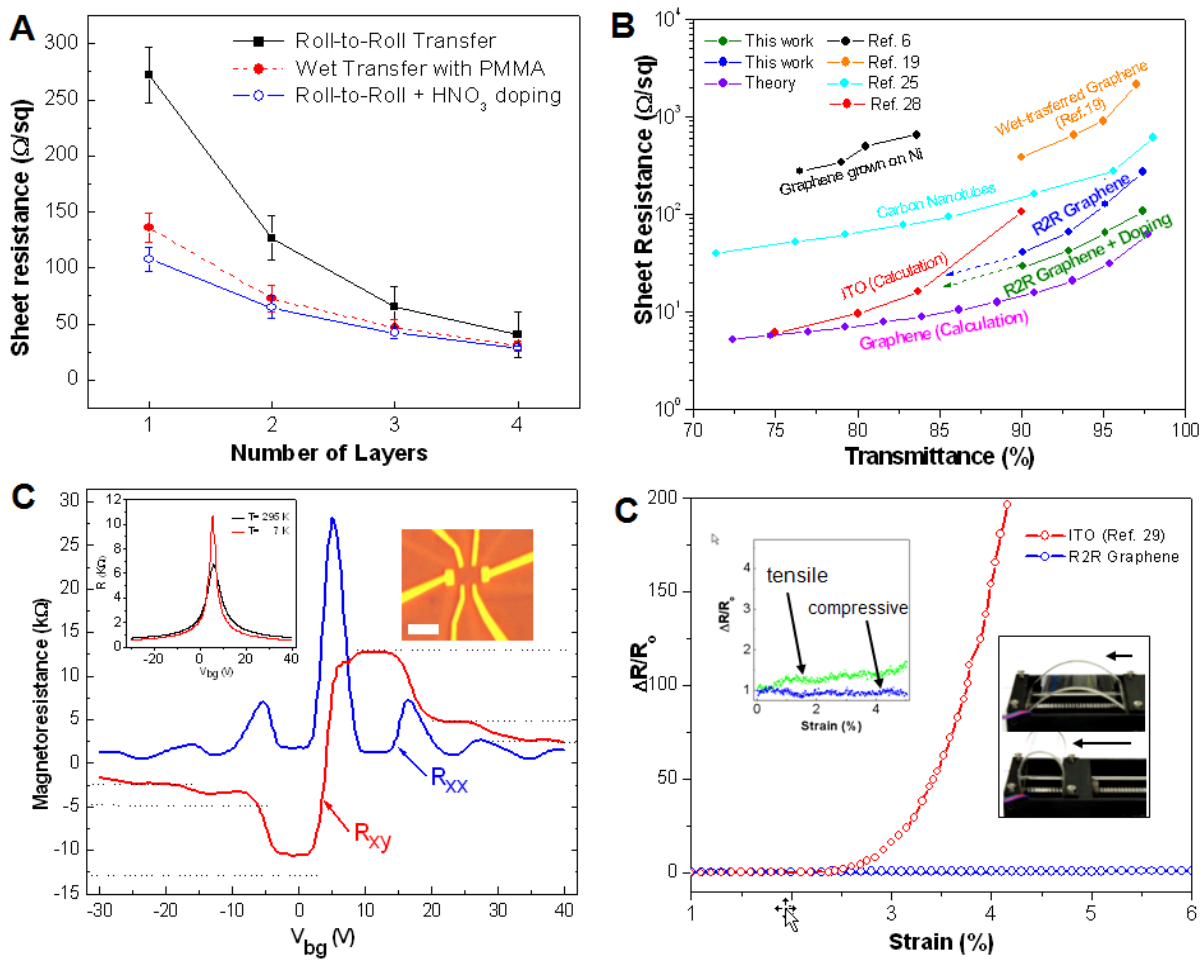

**Fig. 4**. Electrical characterizations of layer-by-layer transferred and HNO$_3$-doped graphene films. (**A**) Sheet resistances of transferred graphene films using a roll-to-roll dry transfer method combined with thermal release tapes and a PMMA-assisted wet transfer method. (**B**) Comparison of sheet resistance vs. transmittance plots from previous reference. The scheme is borrowed from Ref. 20. (**C**) Electrical properties of a monolayer graphene hall bar device. Four-probe resistivity (left bottom insert) is measured as a function of gate voltage in a monolayer graphene Hall bar shown in the insert (right) at room temperature (black curve) and T=6 K (red curve). QHE effect at $T$ = 6K and $B$ = 9T measured in the same device. The longitudinal resistivity $\rho_{xx}$ and Hall conductivity $\sigma_{xy}$ are plotted as a function of gate voltage. The sequence of the first three half-integer plateaus corresponding to ν =2, 6, and 10, typical for single layer graphene are clearly seen. The hall effect mobility of this device is $\mu_{Hall}$ = 7350 cm$^{-2}$/Vs. The scale bar in the insert figure is 3 μm. (**D**) Electromechanical properties of graphene-based touch screen devices compared with ITO/PET electrodes. The inset shows the resistance change by compressive and tensile strain applied to the upper and lower graphene/PET panels, respectively.

# Supporting Material

## 30-Inch Roll-Based Production of High-Quality Graphene Film for Flexible Transparent Electrodes

Sukang Bae[1*], Hyeong Keun Kim[3*], Youngbin Lee[1], Xianfang Xu[5], Jae-Sung Park[7], Yi Zheng[5], Jayakumar Balakrishnan[5], Danho Im[2], Tian Lei[1], Young Il Song[6], Young Jin Kim[1,3], Kwang S. Kim[7], Barbaros Özyilmaz[5], Jong-Hyun Ahn[1,4†], Byung Hee Hong[1,2†] & Sumio Iijima[1,8]

### A. Enlarged grain sizes of Cu foil after annealing/growth

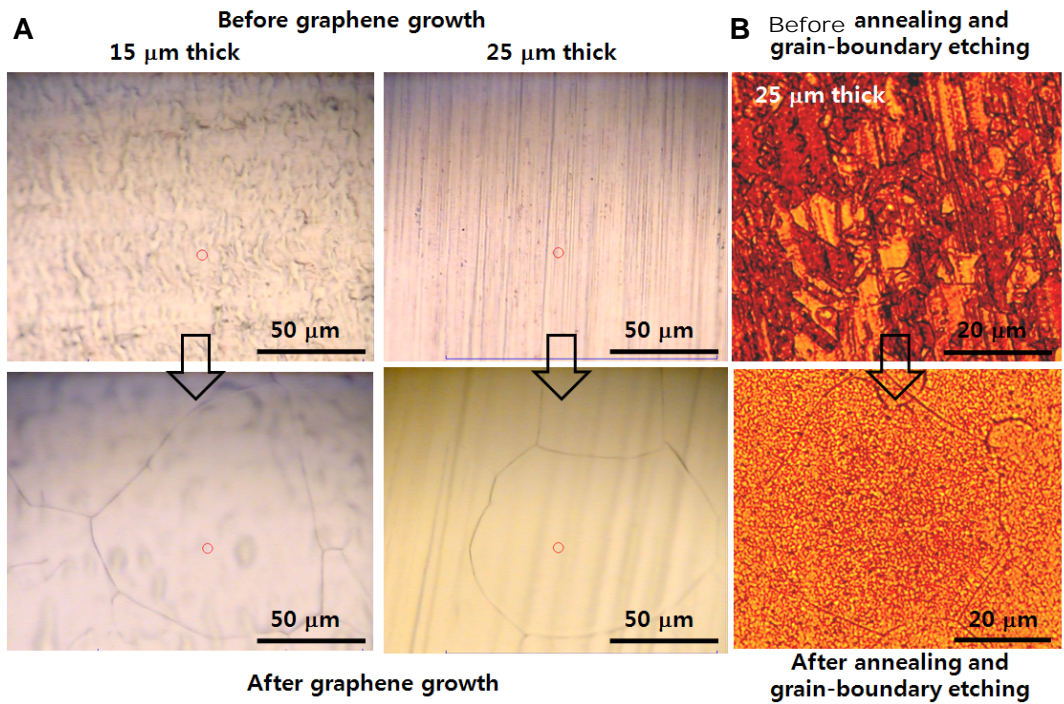

**Fig. S1**. Grain size analyses of Cu foils before and after annealing/growth. (**A**) Optical microscope images of Cu foils before and after graphene growth at 1,000°C. The cracks on foils usually formed at grain boundaries. (**B**) Optical images of polished Cu foils before and after annealing at 1,000°C, followed by brief acid treatment. The grain boundaries are etched faster than single crystalline surfaces, resulting in the formation line patterns on the polished Cu surface.

## B. Homogeneity analysis of graphene films

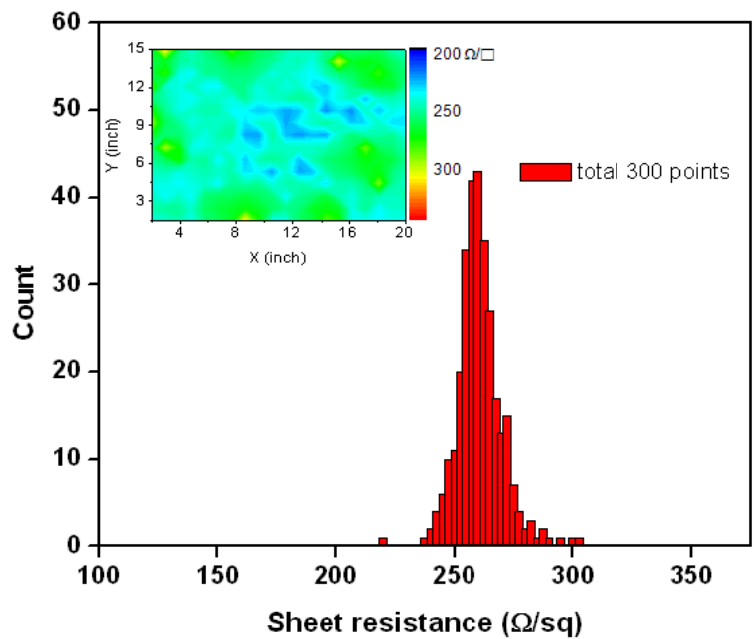

**Fig. S2**. Distribution of sheet resistances tested on the 300 pieces of 1x1 inch$^2$ monolayer graphene/PET films. The inset shows the corresponding spatial distribution on a 20x15 inch$^2$ graphene/PET film.

## C. AFM analysis of graphene films

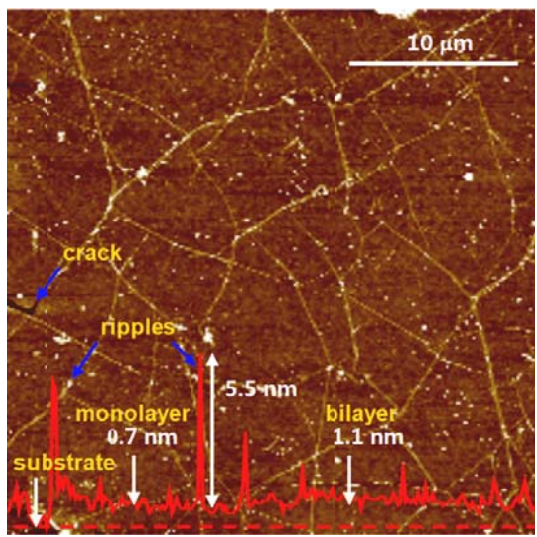

**Fig. S3.** A contact-mode AFM image of the graphene film transferred on a PET film. A few nm high ripples are usually formed due to the different thermal expansion of Cu and graphene, and the cracks are formed during transfer process. The height profile (red solid line) measured along the dashed red line indicates the thicknesses of mono- and bi-layers approximately.

### D. TEM images

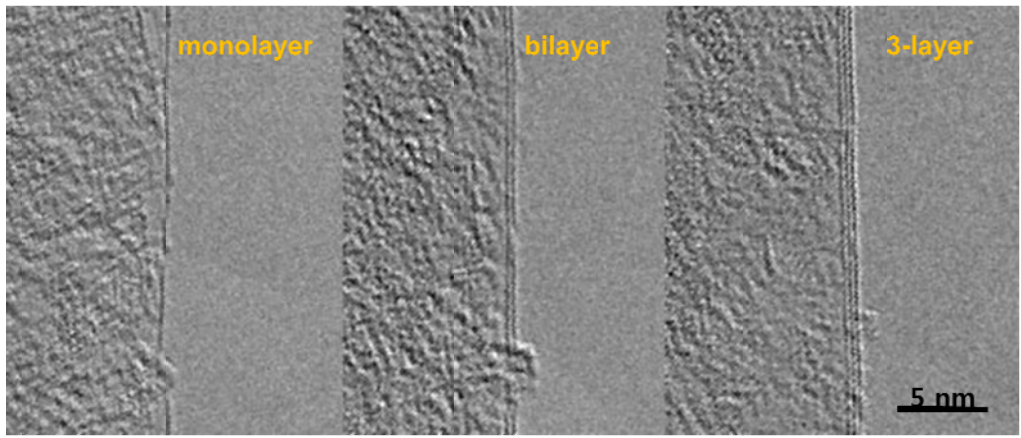

**Fig. S4.** High-resolution TEM images showing different number of graphene layers.

### E. Operation Movie for Graphene-Based Touch Screen Panels

- The movie file will be sent by email upon request. (byunghee@skku.edu)